\documentclass[aps, prd, amsmath, floats, floatfix, twocolumn, nofootinbib, superscriptaddress]{revtex4-1}
\usepackage{graphicx}
\usepackage{color}
\usepackage{latexsym}
\usepackage[colorlinks]{hyperref}

\newcommand{\be}{\begin{eqnarray}}
\newcommand{\ee}{\end{eqnarray}}

\begin{document}

\title{About the Kerr nature of the stellar-mass black hole in GRS~1915+105}

\author{Yuexin~Zhang}
\affiliation{Center for Field Theory and Particle Physics and Department of Physics, Fudan University, 200438 Shanghai, China}

\author{Askar~B.~Abdikamalov}
\affiliation{Center for Field Theory and Particle Physics and Department of Physics, Fudan University, 200438 Shanghai, China}

\author{Dimitry~Ayzenberg}
\affiliation{Center for Field Theory and Particle Physics and Department of Physics, Fudan University, 200438 Shanghai, China}

\author{Cosimo~Bambi}
\email[Corresponding author: ]{bambi@fudan.edu.cn}
\affiliation{Center for Field Theory and Particle Physics and Department of Physics, Fudan University, 200438 Shanghai, China}

\author{Thomas~Dauser}
\affiliation{Remeis Observatory \& ECAP, Universit\"{a}t Erlangen-N\"{u}rnberg, 96049 Bamberg, Germany}

\author{Javier~A.~Garc{\'\i}a}
\affiliation{Cahill Center for Astronomy and Astrophysics, California Institute of Technology, Pasadena, CA 91125, USA}
\affiliation{Remeis Observatory \& ECAP, Universit\"{a}t Erlangen-N\"{u}rnberg, 96049 Bamberg, Germany}

\author{Sourabh~Nampalliwar}
\affiliation{Theoretical Astrophysics, Eberhard-Karls Universit\"at T\"ubingen, 72076 T\"ubingen, Germany}

\begin{abstract}
We employ the accretion disk reflection model {\sc relxill\_nk} to test the spacetime geometry around the stellar-mass black hole in GRS~1915+105. We adopt the Johannsen metric with the deformation parameters $\alpha_{13}$ and $\alpha_{22}$, for which the Kerr solution is recovered when $\alpha_{13} = \alpha_{22} = 0$. We analyze a \textsl{NuSTAR} observation of 2012, obtaining vanishing and non-vanishing values of the deformation parameters depending on the astrophysical model adopted. Similar difficulties were not found in our previous tests with other sources. The results of this work can shed light on the choice of sources suitable for testing the Kerr metric using X-ray reflection spectroscopy and on the parts of our reflection models that more urgently require improvement. 
\end{abstract}

\maketitle


\section{Introduction}

Astrophysical black holes are commonly thought to be the Kerr black holes predicted by Einstein's theory of general relativity~\cite{a1,a2}. Nevertheless, it is important to bear in mind that Einstein's gravity has been extensively tested only in weak gravitational fields~\cite{will}. There are many alternative and modified theories of gravity that have the same predictions as Einstein's gravity in the weak field regime but show deviations from general relativity when gravity becomes strong. Astrophysical black holes are an ideal laboratory for testing Einstein's theory of general relativity in the strong field regime.

Nowadays, there are two lines of research to test black holes and strong gravity: $i)$ the study of the properties of electromagnetic radiation emitted by material orbiting close to a black hole~\cite{r2,r3,r5,r6}, and $ii)$ the study of the gravitational wave signal emitted by black hole systems~\cite{r1,r4,abbott,yyp}. The two approaches are complementary because they probe different sectors of the theory. Electromagnetic techniques can, strictly speaking, only test the motion of massless and massive particles in the strong gravity region. The gravitational wave spectrum is sensitive to the evolution of the spacetime metric and thus can test the strong-field, highly-dynamical regime. In the present work, we consider one of the electromagnetic techniques for black hole tests: X-ray reflection spectroscopy.

There are many alternative and modified theories of gravity that have the same predictions as Einstein's gravity in the weak field regime and have black holes different from those of general relativity. Testing the Kerr nature of astrophysical black holes is an important check to confirm the validity of Einstein's gravity in the strong field regime~\cite{r1,r2,r3,r4,r5,r6}.

X-ray reflection spectroscopy is potentially a powerful technique to test the Kerr nature of astrophysical black holes with electromagnetic radiation~\cite{i1,i2,i3,i4,i5,i6,i7,shenzhen}. This technique is based on the study of the reflection spectrum of accretion disks~\cite{k1,k2}. The accretion disk of black holes emits thermal photons that can have inverse Compton scattering off free electrons in the so-called ``corona'', which is a hot, usually compact and optically thin, medium close to the compact object. A fraction of the Comptonized photons illuminate the disk, producing a reflection spectrum with some emission lines. The most prominent features in the reflection spectrum are usually the iron K$\alpha$ line around 6~keV and the Compton hump at 10-30~keV. The observed reflection spectrum, and the iron K$\alpha$ line in particular, are strongly affected by relativistic effects occurring in the strong gravity region around the black hole. In the presence of the correct astrophysical model and high quality data, we can study the features of the reflection spectrum and test the nature of the compact object.

Recently, we have developed the reflection model {\sc relxill\_nk} to probe the spacetime metric around astrophysical black holes and test the Kerr black hole hypothesis using X-ray reflection spectroscopy~\cite{noi0,noi0b}. {\sc relxill\_nk} is the natural extension of the {\sc relxill} model~\cite{ref1,ref2} to non-Kerr spacetimes. In {\sc relxill\_nk}, the spacetime is described by a parametric black hole metric in which a set of ``deformation parameters'' is introduced to quantify possible deviations from the Kerr solution. By comparing X-ray data of astrophysical black holes with the theoretical predictions of {\sc relxill\_nk} we can measure the values of these deformation parameters and check whether they vanish, as is required by Einstein's theory.

In the past year, we have analyzed a few sources with {\sc relxill\_nk}. In the case of supermassive black holes, we have tested 1H0707--495 with \textsl{XMM-Newton} and \textsl{NuSTAR} data~\cite{noi1}, Ark~564 and Mrk~335 with \textsl{Suzaku} data~\cite{noi2,noi5}, and MCG--6--30--15 with combined data of \textsl{XMM-Newton} and \textsl{NuSTAR}~\cite{noi6}. In Ref.~\cite{noi7}, we have presented the analysis of \textsl{Suzaku} data of seven ``bare'' active galactic nuclei (Ton~S180, RBS~1124, Ark~120, Swift~J0501.9--3239, 1H0419--577, PKS~0558--504, and Fairall~9), i.e. sources with no complicating intrinsic absorption. In the case of stellar-mass black holes, we have tested GX~339--4 with \textsl{RXTE} data~\cite{noi3} and GS~1354--645 with \textsl{NuSTAR} data~\cite{noi4}. In all these studies we have found that the measurements of the value of the deformation parameters are consistent with zero at 1- or 2-$\sigma$; that is, our results are consistent with the hypothesis that the spacetime metric around all these objects is described by the Kerr solution within the statistical uncertainties of our measurements; systematic uncertainties are more difficult to estimate and work is underway. The constraints obtained from MCG--6--30--15, GS~1354--645, and some bare active galactic nuclei appear quite stringent and we have shown how imposing unjustified {\it ad hoc} emissivity profiles would completely spoil our results, suggesting that our current theoretical model is good enough to test Einstein's gravity with these sources~\cite{noi4}.

Here we continue our program of testing the Kerr black hole hypothesis with {\sc relxill\_nk} and we present the study of a new source, GRS~1915+105, which is a binary system of a stellar-mass black hole with a low mass companion star. We analyze a \textsl{NuSTAR} observation of 2012, hoping to get strong constraints on the deformation parameters, in analogy with what was obtained for GS~1354--645. Like the latter, GRS~1915+105 has properties that are supposed to help get good constraints on its strong gravity region: its spin parameter is high, so the inner edge of the disk can be very close to the black hole, and the viewing angle is relatively high as well, thus maximizing the relativistic effects of Doppler boosting and light bending. Since it is a stellar-mass black hole, the source is bright and we have a good statistics. \textsl{NuSTAR} data are also suitable for this kind of test, as we can measure the spectrum up to 80~keV and there is no pile-up problem. However, we meet a problem in recovering the Kerr metric. More specifically, we find non-vanishing deformation parameters when we employ the model adopted in~\cite{nustar}, where the authors study this \textsl{NuSTAR} observation assuming the Kerr metric. We try to change the intensity profile, but we constantly do not recover the Kerr solution. When we add a non-relativistic reflection component, we recover Kerr when the intensity profile is modeled with a power law and we do not recover Kerr with a broken power law. In our previous tests, we had never met similar difficulties. We compare the results of this work with those of previous tests of the Kerr metric, and we discuss the differences between GRS~1915+105 and the other sources.

The content of the paper is as follows. In Section~\ref{s-xrs}, we review our method to test the Kerr black hole hypothesis with the reflection model {\sc relxill\_nk}. In Section~\ref{s-obs}, we present the observation and how we reduced the data. Section~\ref{s-ana} is devoted to the data analysis and we show the best-fit values and the constraints on the deformation parameters. In Section~\ref{s-dis}, we discuss our results and we compare them with those obtained in other studies. Throughout the paper we employ a metric with signature $(-+++)$ and units in which $G_{\rm N} = c = 1$.


\section{Testing the Kerr hypothesis with {\sc relxill\_nk} \label{s-xrs}}

The reflection spectrum of accretion disks around black holes originates from the illumination of the accretion disk by Comptonized photons from the corona. From the comparison of the theoretical predictions with observational data, it is possible to infer the properties of the system. Our disk's reflection model for non-Kerr spacetimes is called {\sc relxill\_nk} and was presented in Refs.~\cite{noi0,noi0b}. It is the natural extension of the {\sc relxill} model~\cite{ref1,ref2}, in which the background metric is assumed to be described by the Kerr solution. {\sc relxill} itself is the result of the merger of two models: {\sc xillver} and {\sc relconv}. {\sc xillver} is a purely atomic model to calculate the reflection spectrum in the rest-frame of the gas of the accretion disk~\cite{x1,x2}. {\sc relconv} is a convolution model and transforms the reflection spectrum calculated by {\sc xillver} into that detected far from the source taking all relativistic effects (Doppler boosting, gravitational redshift, and light bending) into account~\cite{x3}. In {\sc relxill\_nk} we have extended the convolution model {\sc relconv} in order to calculate the detected spectrum in the case of a non-Kerr spacetime.

In what follows, we assume that the geometry of the spacetime is described by the Johannsen metric~\cite{tj}\footnote{The Johannsen metric is not a black hole solution of any modified theory of gravity but simply a parametric black hole metric aiming at describing a spacetime with possible deviations from the Kerr solution. Here we employ the Johannsen metric with the spirit to perform a null experiment and check whether astrophysical observations require that the value of the deformation parameters is consistent with zero as requested by the Kerr hypothesis.}. In Boyer-Lindquist-like coordinates, the line element reads
\be\label{eq-jm}
ds^2 &=&-\frac{\tilde{\Sigma}\left(\Delta-a^2A_2^2\sin^2\theta\right)}{B^2}dt^2
+\frac{\tilde{\Sigma}}{\Delta}dr^2+\tilde{\Sigma} d\theta^2 \nonumber\\
&&-\frac{2a\left[\left(r^2+a^2\right)A_1A_2-\Delta\right]\tilde{\Sigma}\sin^2\theta}{B^2}dtd\phi \nonumber\\
&&+\frac{\left[\left(r^2+a^2\right)^2A_1^2-a^2\Delta\sin^2\theta\right]\tilde{\Sigma}\sin^2\theta}{B^2}d\phi^2
\ee
where $M$ is the black hole mass, $a = J/M$, $J$ is the black hole spin angular momentum, $\tilde{\Sigma} = \Sigma = f$, and
\be
&& \Sigma = r^2 + a^2 \cos^2\theta \, , \qquad
\Delta = r^2 - 2 M r + a^2 \, , \nonumber\\
&& B = \left(r^2+a^2\right)A_1-a^2A_2\sin^2\theta \, .
\ee
The functions $f$, $A_1$, $A_2$, and $A_5$ are defined as
\be
f &=& \sum^\infty_{n=3} \epsilon_n \frac{M^n}{r^{n-2}} \, , \\
A_1 &=& 1 + \sum^\infty_{n=3} \alpha_{1n} \left(\frac{M}{r}\right)^n \, , \\
A_2 &=& 1 + \sum^\infty_{n=2} \alpha_{2n}\left(\frac{M}{r}\right)^n \, , \\
A_5 &=& 1 + \sum^\infty_{n=2} \alpha_{5n}\left(\frac{M}{r}\right)^n \, .
\ee
$\{ \epsilon_n \}$, $\{ \alpha_{1n} \}$, $\{ \alpha_{2n} \}$, and $\{ \alpha_{5n} \}$ are four infinite sets of deformation parameters without constraints from the Newtonian limit and weak field experiments, and the Kerr metric is recovered when all deformation parameters vanish. In this paper, we will only focus on the deformation parameters $\alpha_{13}$ and $\alpha_{22}$, as they are the two with the strongest impact on the reflection spectrum~\cite{noi0}. In what follows, we will consider the possibility that one of the two deformation parameters may be non-vanishing and we will try to infer its value from the data of GRS~1915+105. First, we will try to measure $\alpha_{13}$ assuming that $\alpha_{22}=0$ and then we will consider the opposite case, namely $\alpha_{13}=0$ and we try to measure the value of $\alpha_{22}$. The possibility of two variable deformation parameters at the same time is beyond the capabilities of our current version of {\sc relxill\_nk}.

Note that, in order to avoid spacetimes with pathological properties, we have to impose some restrictions on the values of the spin parameter $a_* = a/M$ and of the deformation parameters $\alpha_{13}$ and $\alpha_{22}$. As in the case of the Kerr spacetime, we require that $| a_* | \le 1$, because for $| a_* | > 1$ there is no black hole but a naked singularity. As discussed in~\cite{tj,noi2}, we also have to impose the following restrictions on $\alpha_{13}$ and $\alpha_{22}$
\be
\label{eq-constraints}
&& \alpha_{13} > - \frac{1}{2} \left( 1 + \sqrt{1 - a^2_*} \right)^4 \, , \\
&& - \left(1 + \sqrt{1 - a_*^2} \right)^2 < \alpha_{22} < \frac{\left( 1 + \sqrt{1 - a^2_*} \right)^4}{a_*^2} \, .
\label{eq-constraints2}
\ee

\begin{figure*}[t]
\begin{center}
\vspace{0.3cm}
\includegraphics[width=0.46\textwidth]{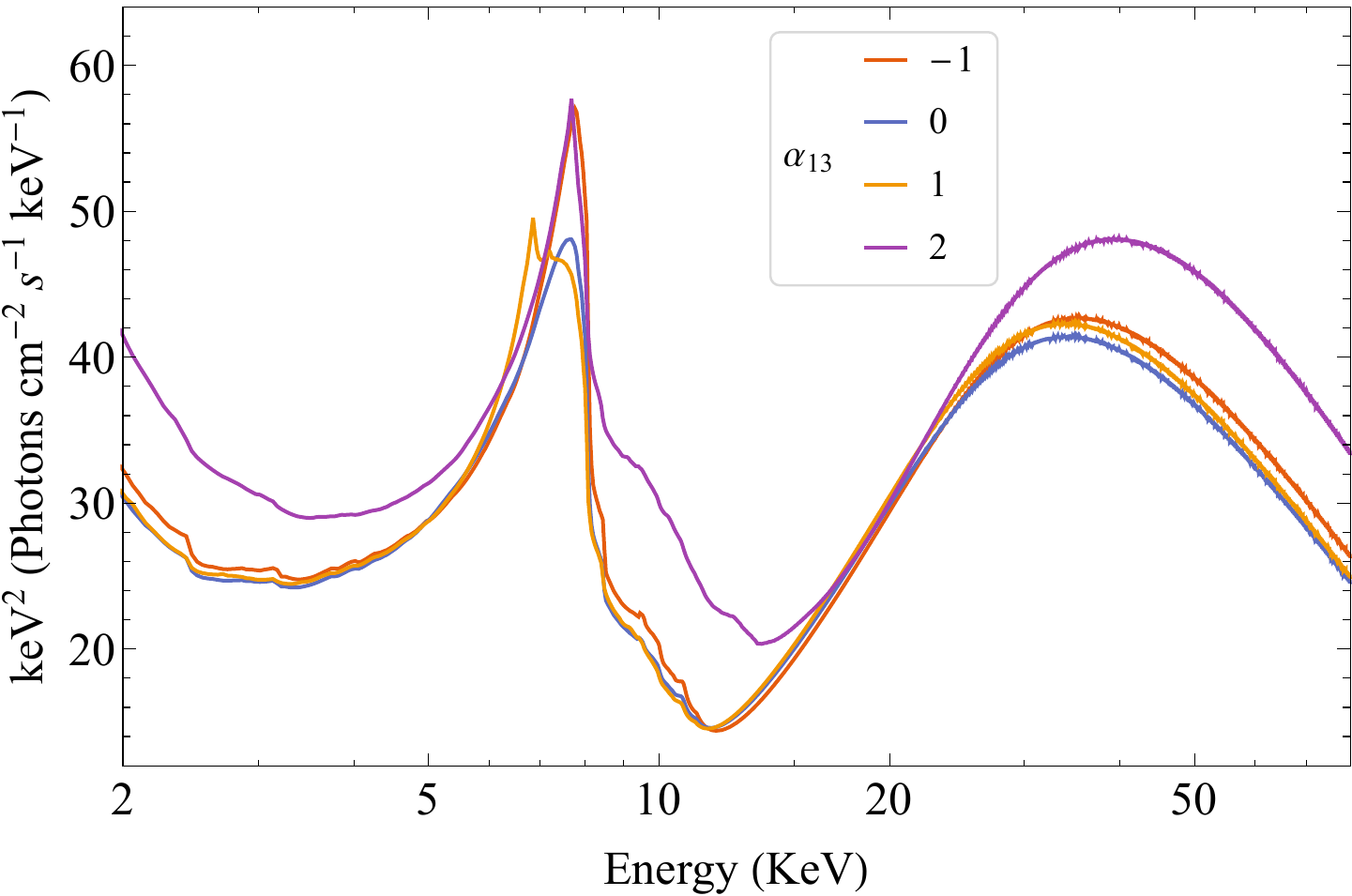}
\hspace{0.8cm}
\includegraphics[width=0.46\textwidth]{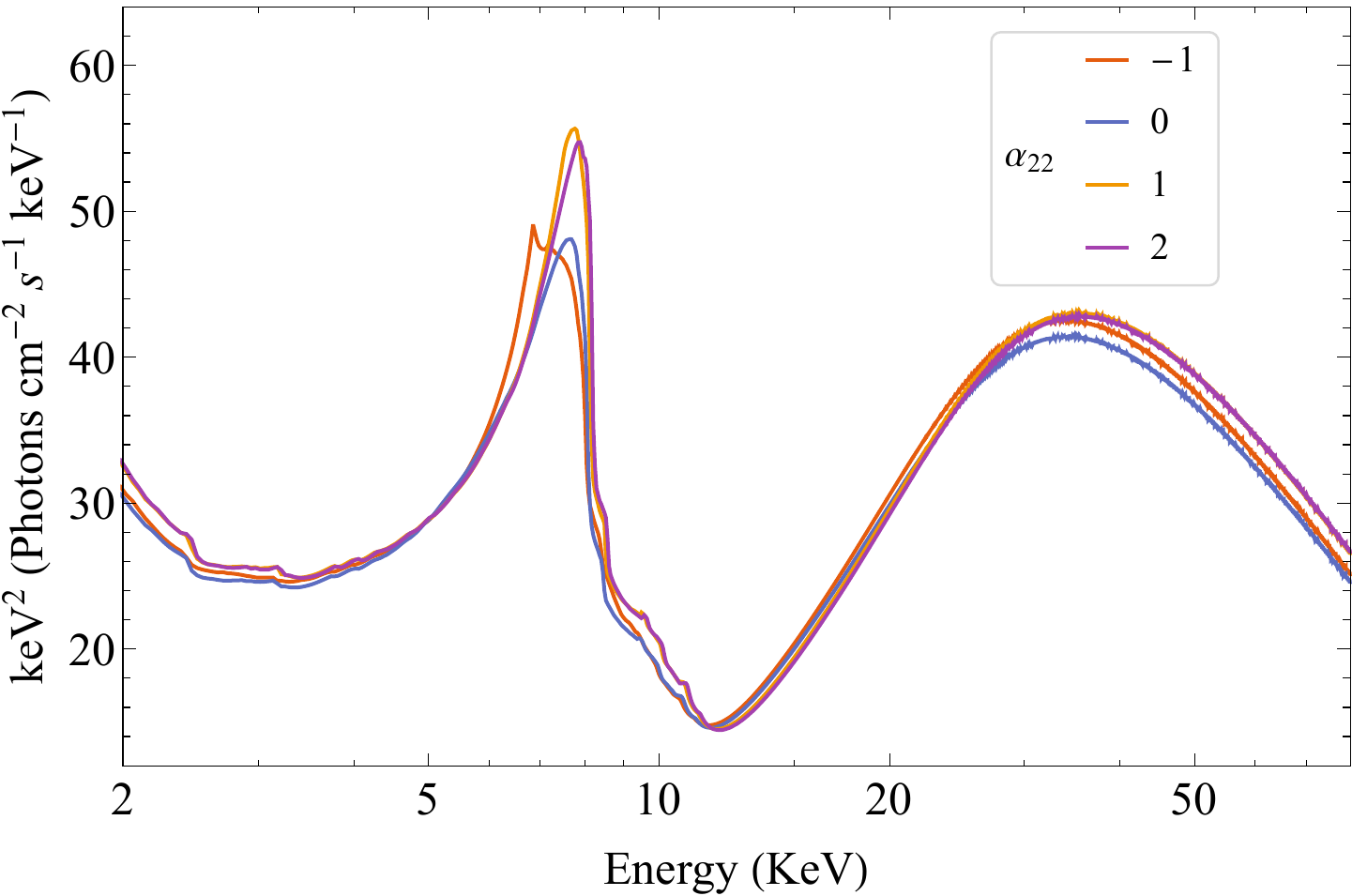}
\end{center}
\vspace{-0.4cm}
\caption{Impact of the deformation parameters $\alpha_{13}$ and $\alpha_{22}$ on the reflection spectrum detected far from the source. In the left panel, $\alpha_{13} = -1, 0, 1, 2$ and $\alpha_{22} =0$. In the right panels, $\alpha_{13}=0$ and $\alpha_{22} = -1, 0, 1, 2$. The values of the other parameters are: $a_* = 0.97$, $q=3$, $\log\xi = 3.1$ ($\xi$ in units erg~cm~s$^{-1}$), $A_{\rm Fe} = 5$, and $i = 60^\circ$. \label{f-alpha}}
\end{figure*}

More details on the astrophysical model employed in {\sc relxill\_nk} can be found in~\cite{noi0,noi0b}. Here we just remind the reader that the accretion disk is assumed to be infinitesimally thin and on the equatorial plane; i.e., orthogonal to the black hole spin. The gas of the accretion disk follows nearly geodesic, equatorial, circular orbits. For thin accretion disk, the inner edge can be at or outside the innermost stable circular orbit (or ISCO); in the present work, we make the standard assumption to fix it at the ISCO radius. The emissivity profile of the accretion disk can be modeled with a power law (i.e. the intensity is proportional to $1/r^q$, where $q$ is some emissivity index) or a broken power law (i.e. the intensity is proportional to $1/r^{q_{\rm in}}$ for $r < R_{\rm br}$ and to $1/r^{q_{\rm out}}$ for $r > R_{\rm br}$, where $q_{\rm in}$ and $q_{\rm out}$ are the inner and the outer emissivity index, respectively, and $R_{\rm br}$ is called the breaking radius). The ionization of the accretion disk is described by a single ionization parameter $\xi$ and the composition of the accretion disk is taken into account by the iron abundance $A_{\rm Fe}$. An important parameter in the model is the viewing angle $i$, namely the angle between our line of sight and the spin of the black hole. Fig.~\ref{f-alpha} shows the typical reflection spectrum as calculated by {\sc relxill\_nk} for a few values of $\alpha_{13}$ and $\alpha_{22}$, respectively. Note that the reflection spectrum of the disk does not directly depend on the black hole mass (the mass, however, regulates the temperature of the disk and, in turn, the ionization parameter $\xi$). Therefore, unlike the continuum-fitting method, X-ray reflection spectroscopy does not need any independent measurement of the black hole mass to fit the spectrum of the source.

\section{Observations and data reduction \label{s-obs}}

GRS~1915+105 is quite a special source. While it is a low mass X-ray binary (i.e. the mass of the companion star is less than a few Solar masses), it is a persistent X-ray source since 1992. This is probably due to its large accretion disk, which is capable of providing a sufficiently high mass transfer at any time.

\textsl{NuSTAR} observed GRS~1915+105 on 2012 July 3 for approximately 60~ks. This observation was analyzed for the first time in~\cite{nustar}, where the authors -- assuming the Kerr metric -- measured a spin parameter $a_* = 0.98 \pm 0.01$ at 1-$\sigma$ statistical error.

In our analysis, we employ Xspec v12.10.0~\cite{arnaud}. We process the data from both the FPMA and FPMB instruments using \texttt{nupipeline} v0.4.3 with the standard filtering criteria and the NuSTAR CALDB version 20180419. We use the \texttt{nuproducts} routine to extract source spectra, responses, and background spectra. Source spectra are extracted from a circular region of radius $90''$. Background spectra are extracted from regions of equivalent size on each detector. All spectra are grouped to a minimum of 30 counts before analysis to ensure the validity of the $\chi^2$ fit statistics. After all efficiencies and screening, the net exposure time for the resultant spectra is 14.85~ks for FPMA, and 15.31~ks for FPMB.

Assuming the black hole mass $M_{\rm BH} = ( 10.1 \pm 0.6 )$~$M_\odot$ and distance $D = 11$~kpc~\cite{bhmass}, the accretion luminosity of the black hole is $0.23 \pm 0.04$ in Eddington units. It is thus in the range in which the accretion disk is thought to be well described by the Novikov-Thorne model with the inner edge at the ISCO radius~\cite{thin1,thin2}, which is the model employed in our analysis.

\begin{figure*}[t]
\begin{center}
\vspace{0.3cm}
\includegraphics[width=0.46\textwidth]{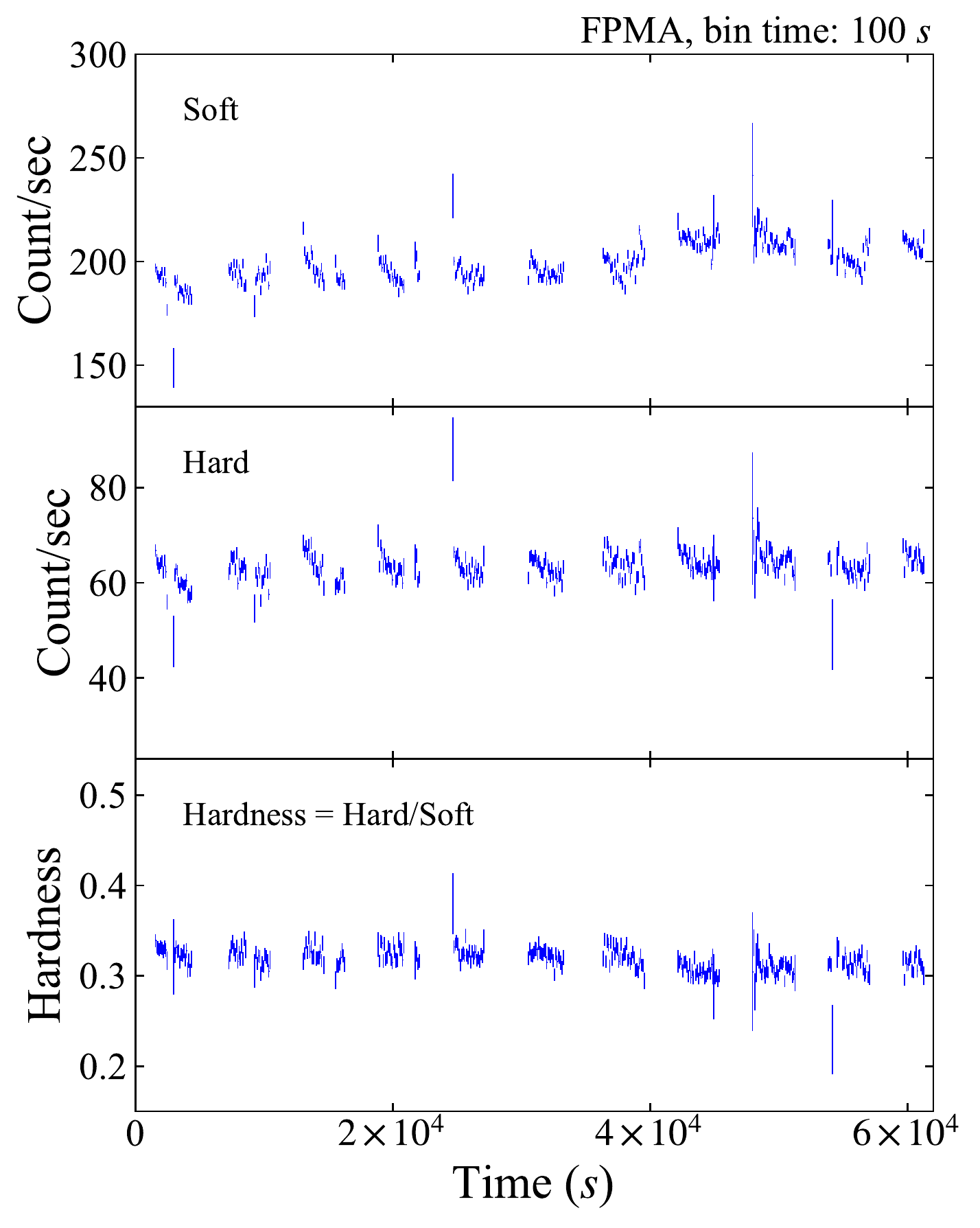}
\hspace{0.5cm}
\includegraphics[width=0.46\textwidth]{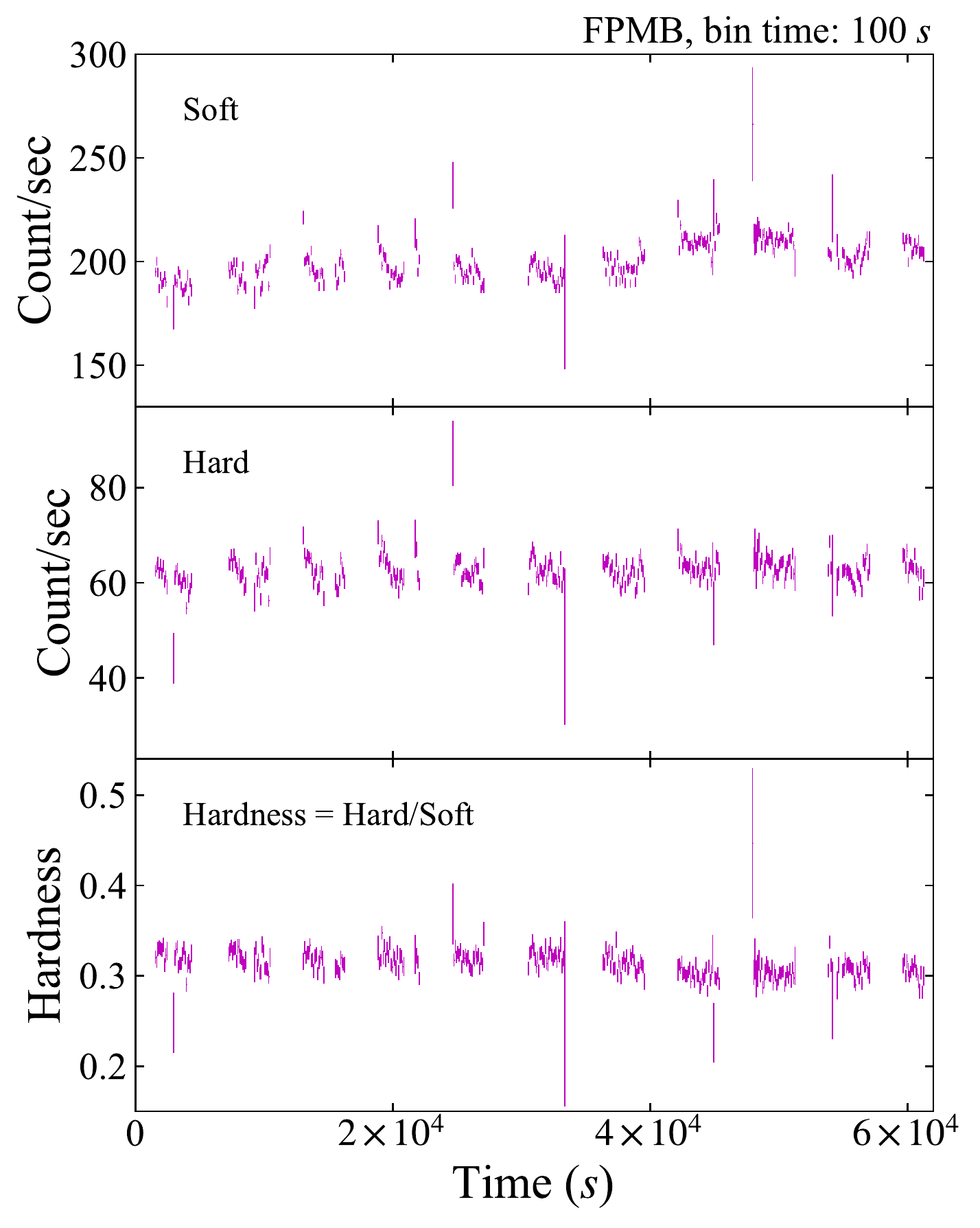}
\end{center}
\vspace{-0.5cm}
\caption{Light curves in the soft (3-10~keV) and hard (10-80~keV) bands of GRS~1915+105 on 2012 July 3 from FPMA (left panel) and FPMB (right panel) and temporal evolution of the hardness of the spectrum. \label{f-lc}}
\end{figure*}

GRS~1915+105 is a highly variable source. However, as shown in Fig.~\ref{f-lc}, the source was quite stable during the 2012 \textsl{NuSTAR} observation and therefore we do not need to take its variability into account in our spectral analysis.

\section{Spectral analysis \label{s-ana}}

\begin{figure*}[t]
\begin{center}
\vspace{0.3cm}
\includegraphics[width=8.5cm]{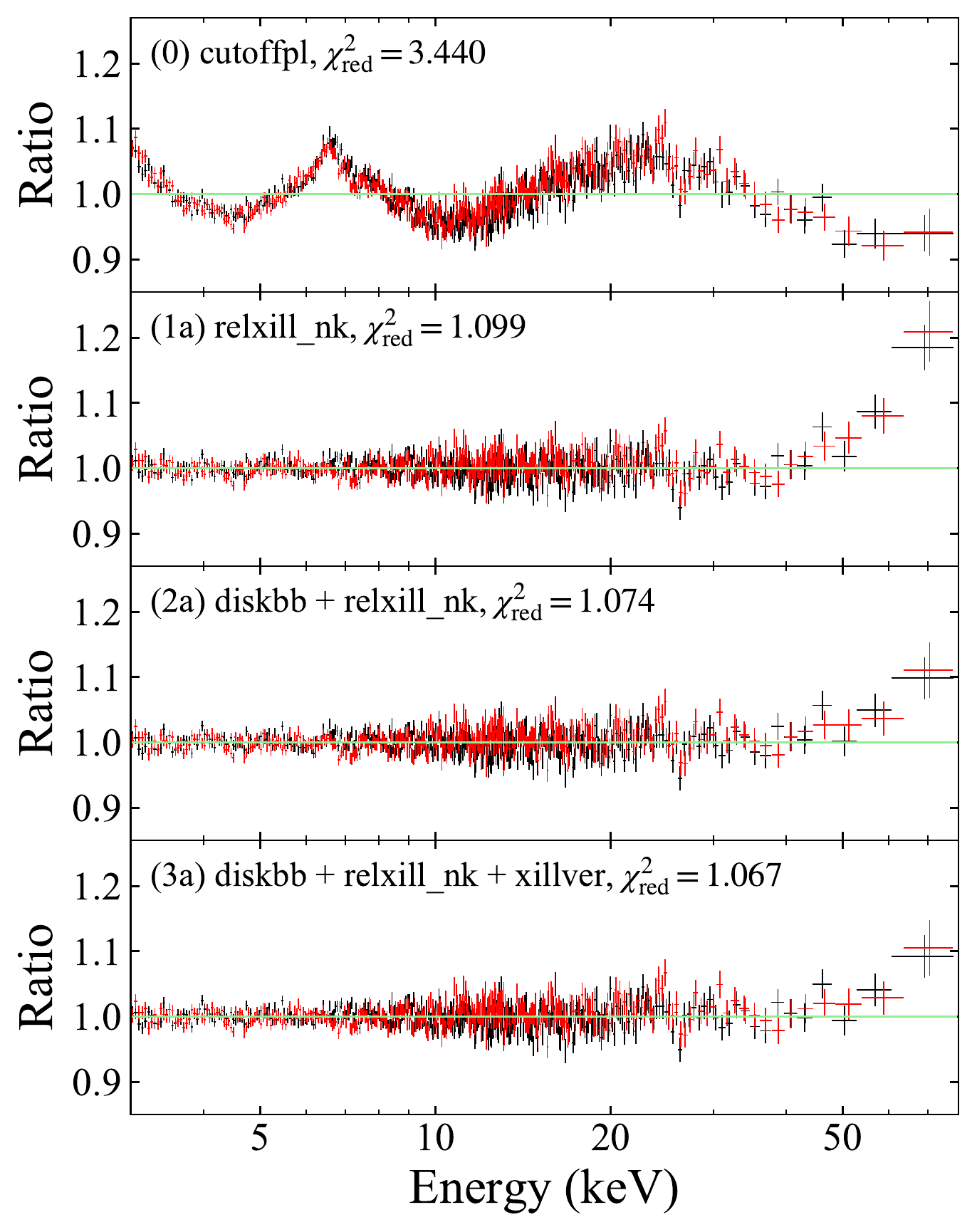}
\hspace{0.2cm}
\includegraphics[width=8.5cm]{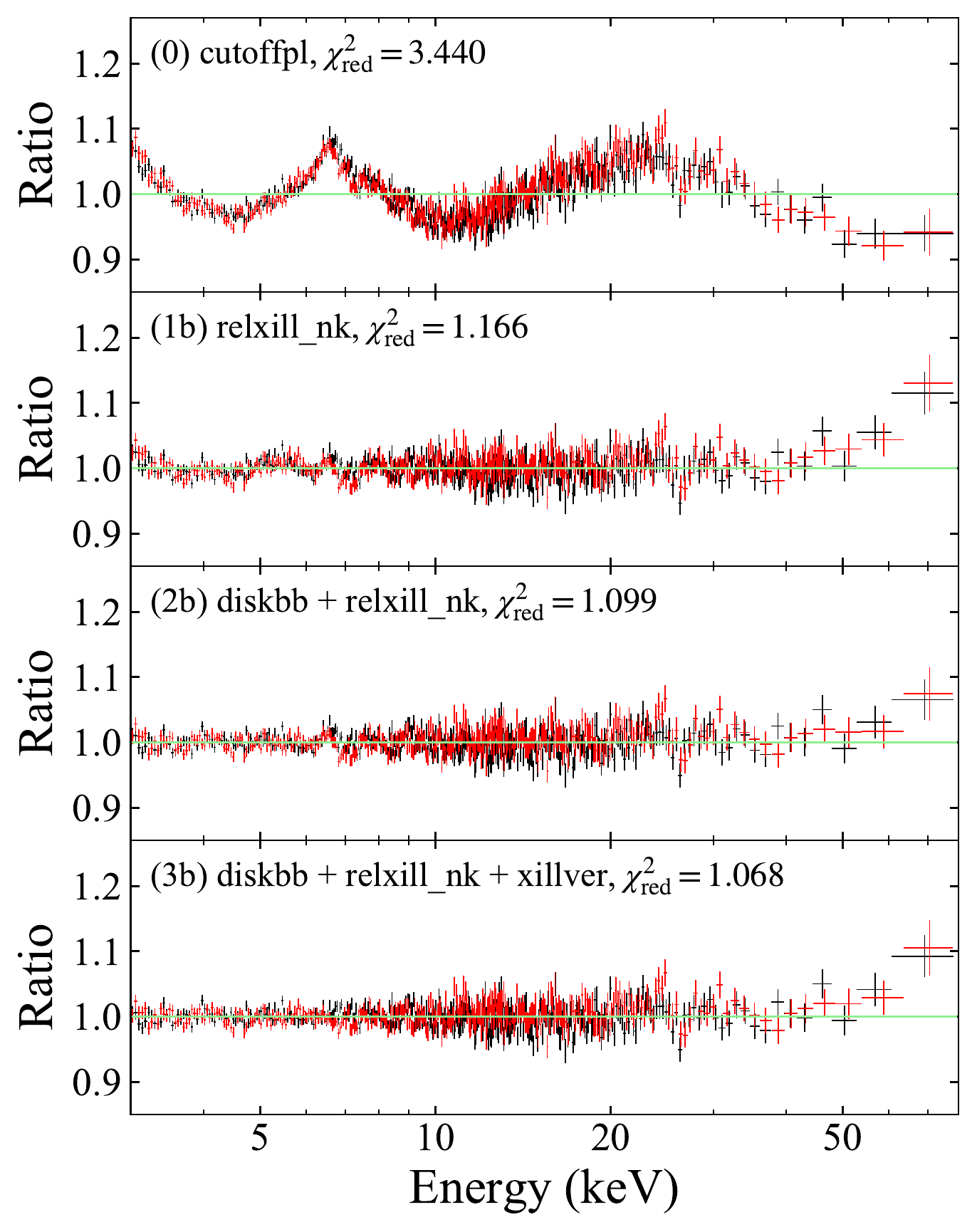}
\end{center}
\vspace{-0.4cm}
\caption{Data to best-fit model ratios for the fits with $\alpha_{13}$ free and $\alpha_{22} =0$ (left panels) and for those with $\alpha_{13}=0$ and $\alpha_{22}$ free (right panels) of the \textsl{NuSTAR} observation of 2012. \label{f-ratio-nustar}}
\end{figure*}

We start fitting the data with a power law component with an exponential cut-off describing the corona spectrum (model~0). The Xspec model is {\sc tbabs*cutoffpl}, where {\sc tbabs} describes the Galactic absorption~\cite{wilms} and {\sc cutoffpl} is for the power law component. The fit is bad and we clearly see a broad iron line around 6~keV and a Compton hump around 20~keV (see the top panels in Fig.~\ref{f-ratio-nustar}).

We improve our model by adding a relativistic reflection component with {\sc relxill\_nk}. Throughout this paper, we employ the version~1.3.2 described in~\cite{noi0b} and available at~\footnote{\tiny http://www.physics.fudan.edu.cn/tps/people/bambi/Site/RELXILL\_NK.html}. We have three models: model~1 in which $\alpha_{13}=\alpha_{22}=0$ (Kerr spacetime), model~1$a$ in which $\alpha_{13}$ free and $\alpha_{22}=0$, and model~1$b$ with $\alpha_{13}=0$ and $\alpha_{22}$ free. Fig.~\ref{f-ratio-nustar} shows the ratio plots of models~1$a$ and 1$b$, where we can clearly see that the fits are significantly better than model~0. We still have an excess of counts at low (around 3~keV) and high (above 40~keV) energies.

\begin{figure*}[t]
\begin{center}
\vspace{0.3cm}
\includegraphics[width=8.5cm,trim={1cm 0 0 0},clip]{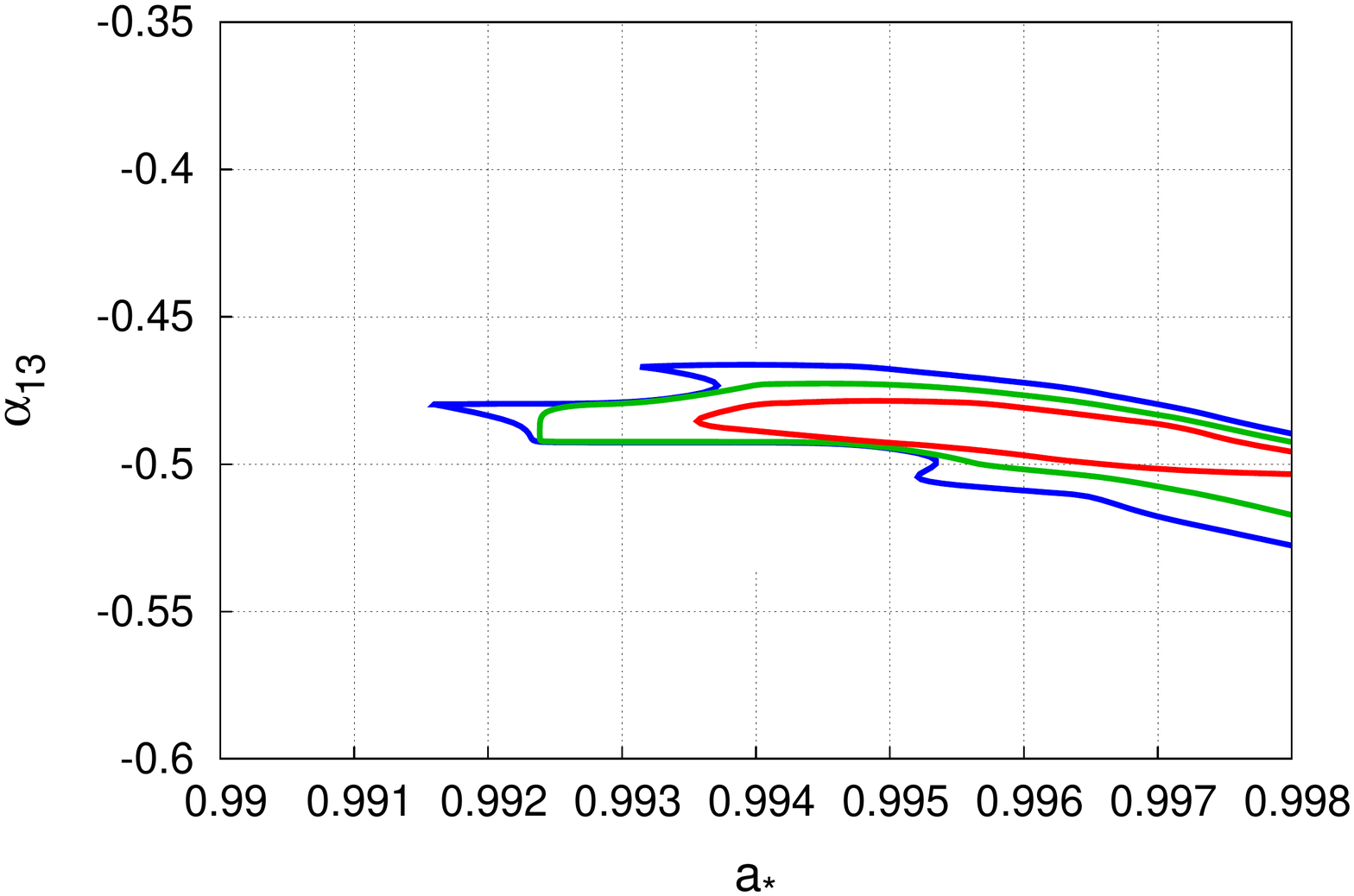}
\includegraphics[width=8.5cm,trim={1cm 0 0 0},clip]{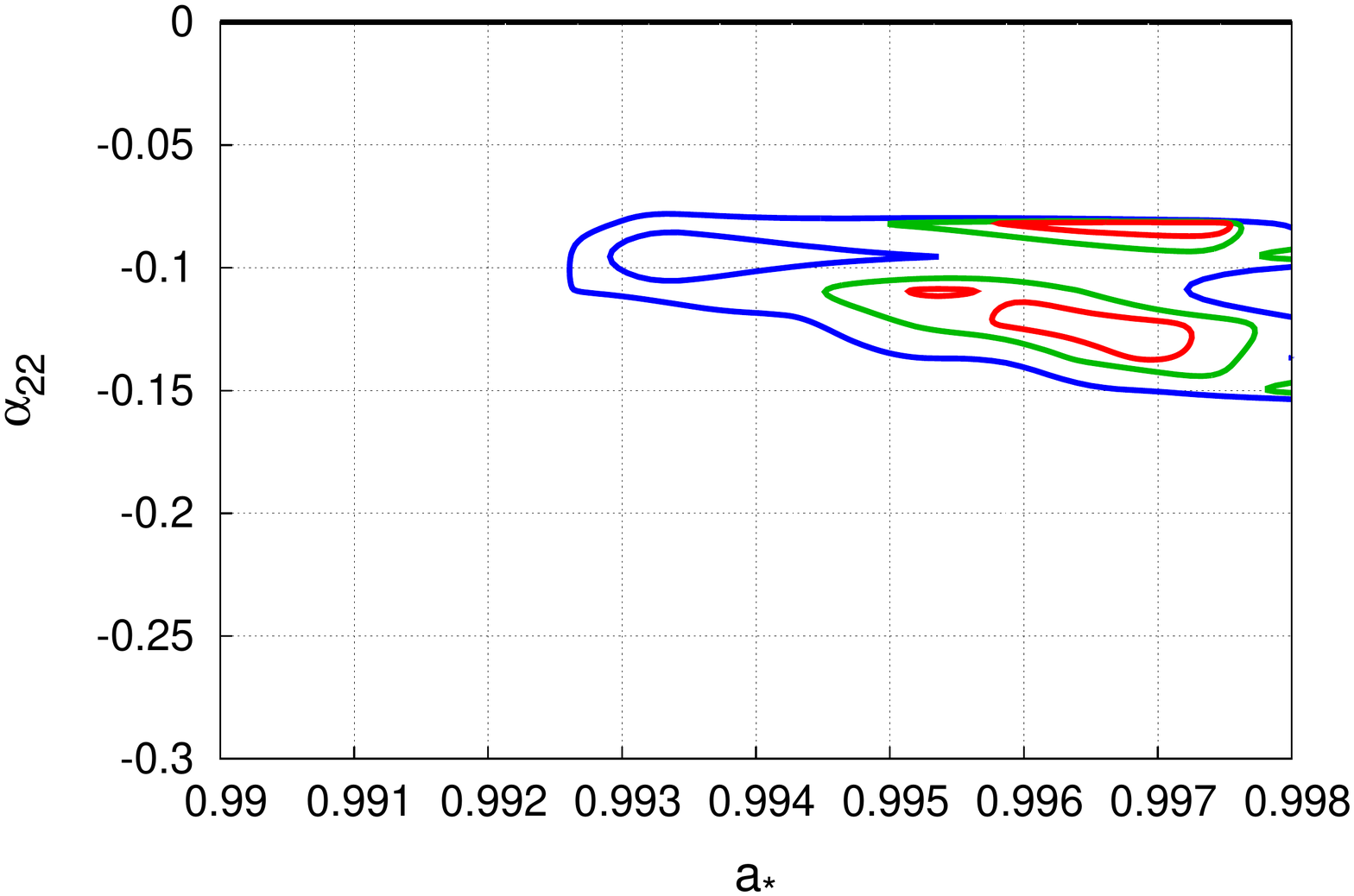}
\end{center}
\vspace{-1.3cm}
\caption{Left panel: Constraints on the spin parameter $a_*$ and on the Johannsen deformation parameter $\alpha_{13}$ according to model~2$a$. Right panel: Constraints on the spin parameter $a_*$ and on the Johannsen deformation parameter $\alpha_{22}$ according to model~2$b$. The red, green, and blue lines indicate, respectively, the 68\%, 90\%, and 99\% confidence level contours for two relevant parameters. \label{f-plots-nustar}}
\vspace{-0.3cm}
\begin{center}
\vspace{0.3cm}
\includegraphics[width=8.5cm,trim={1cm 0 0 0},clip]{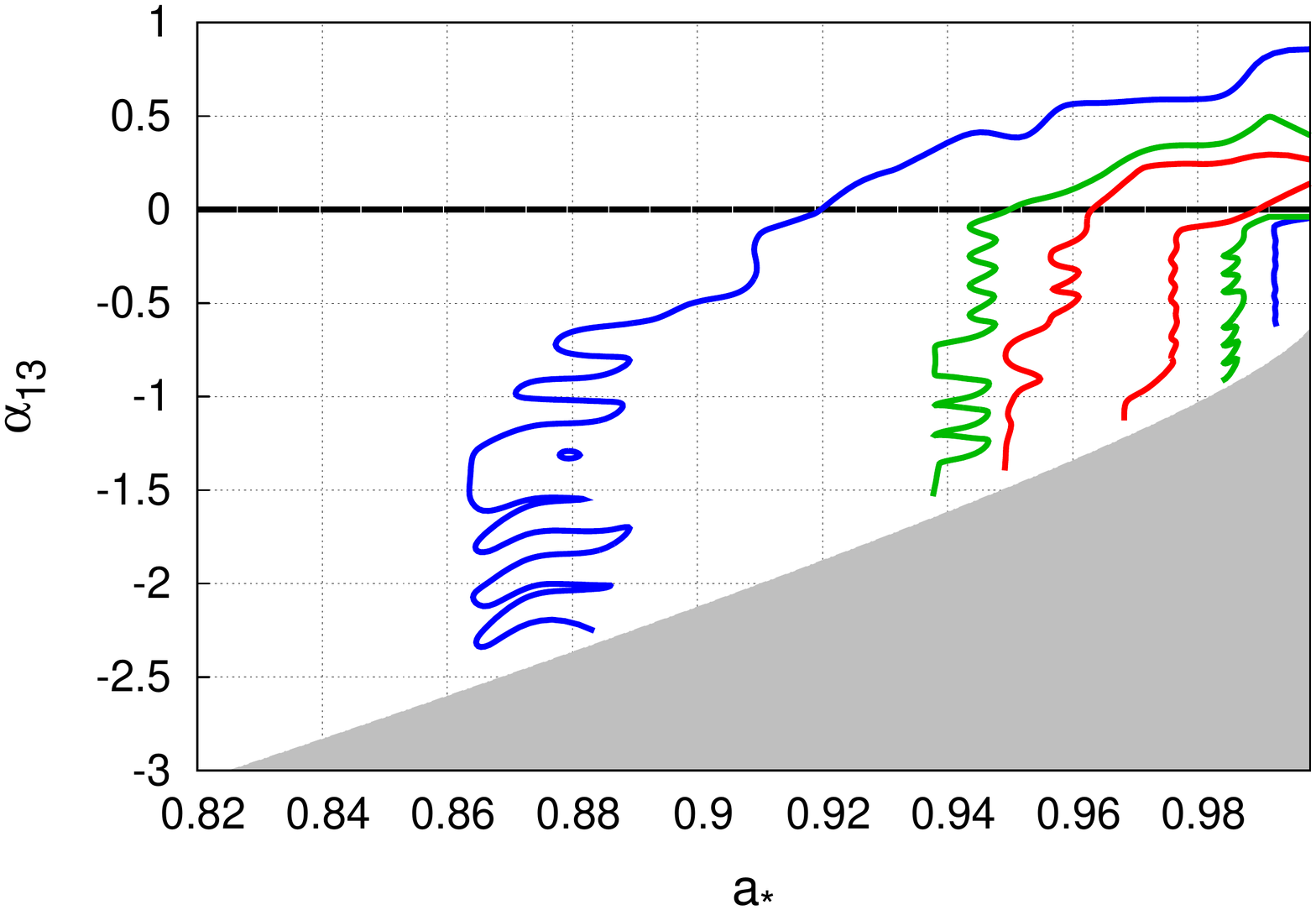}
\includegraphics[width=8.5cm,trim={1cm 0 0 0},clip]{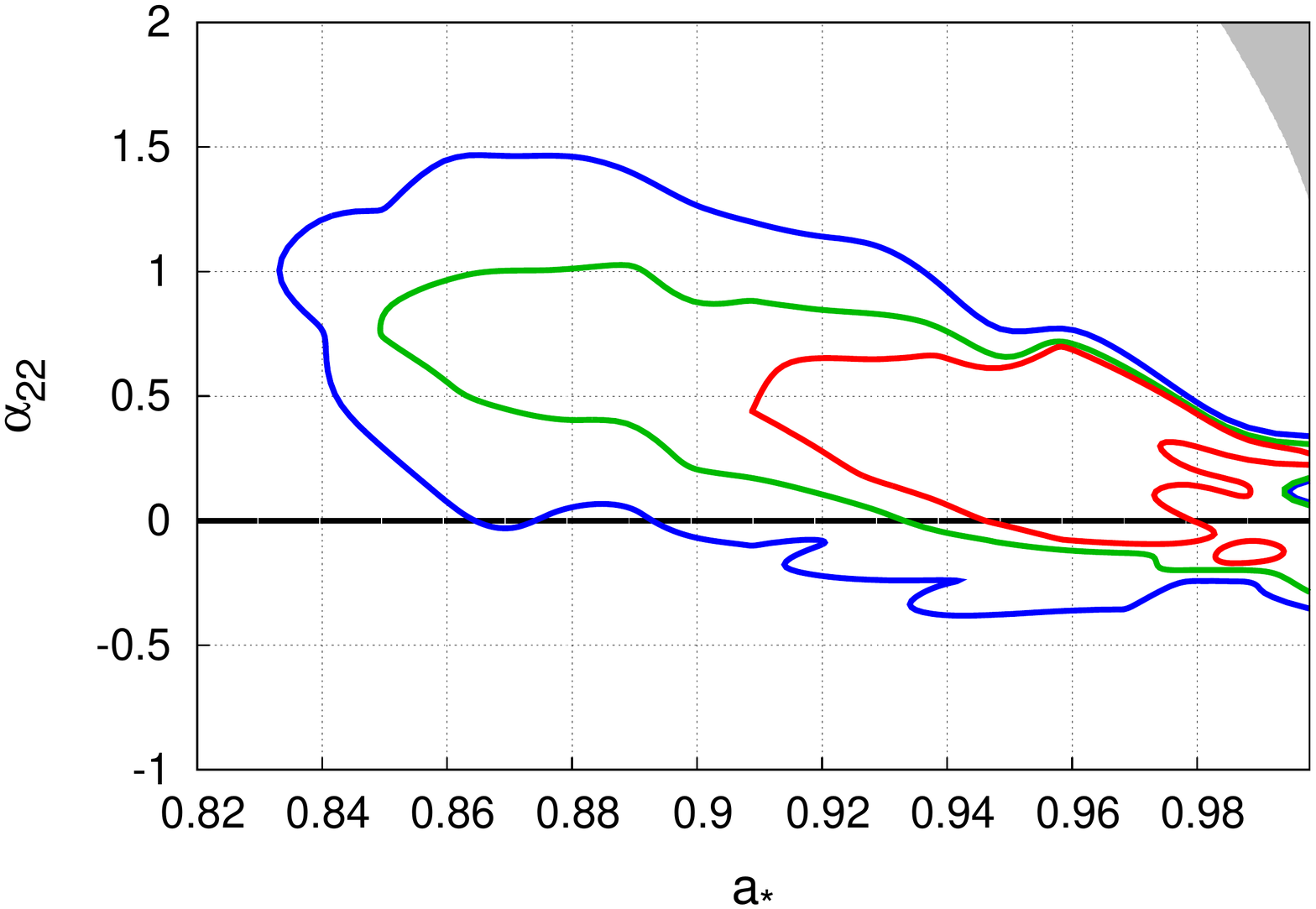}
\end{center}
\vspace{-1.3cm}
\caption{Left panel: Constraints on the spin parameter $a_*$ and on the Johannsen deformation parameter $\alpha_{13}$ according to model~3$a$. Right panel: Constraints on the spin parameter $a_*$ and on the Johannsen deformation parameter $\alpha_{22}$ according to model~3$b$. The red, green, and blue lines indicate, respectively, the 68\%, 90\%, and 99\% confidence level contours for two relevant parameters. The solid black line marks the Kerr solution. The grayed regions are ignored in our study because they do not meet the conditions in Eqs.~(\ref{eq-constraints}) and (\ref{eq-constraints2}). \label{f-plots2-nustar}}
\end{figure*}

We add a thermal component for the accretion disk, which is often present in the spectrum of GRS~1915+105. Again, we have three variants: model~2 in which we assume the Kerr metric, model~2$a$ in which $\alpha_{13}$ is free and $\alpha_{22}=0$, and model~2$b$ with $\alpha_{13}=0$ and $\alpha_{22}$ is free. For the disk's thermal spectrum, we use the Xspec model {\sc diskbb}~\cite{diskbb}, so the total model is {\sc tbabs*(diskbb + relxill\_nk)}. Tab.~\ref{t-fit-kerr} shows the best-fit values for model~2 and Tab.~\ref{t-fit-nustar} does the same for models~2$a$ and 2$b$. The data to best-fit model ratios for models~2$a$ and 2$b$ are reported in Fig.~\ref{f-ratio-nustar} and we can see that {\sc diskbb} improves the quality of the fit. However, in models~2$a$ and 2$b$ we do not recover the Kerr solution. In particular, the difference of $\chi^2$ between the Kerr model and model~2$a$ is $\Delta\chi^2 = 66$. The degeneracy between the spin and the deformation parameters $\alpha_{13}$ and $\alpha_{22}$ is shown in Fig.~\ref{f-plots-nustar}. As discussed in~\cite{noi4}, the intensity profile can play an important role in the estimate of the deformation parameters. We thus try to recover the Kerr solution by fitting the data with various intensity profiles: power law, broken power law with two free emissivity indices, and broken power law with free inner emissivity index and outer emissivity index frozen to 3. The measurement of the deformation parameters can somewhat change but, especially in the case of $\alpha_{13}$, it remains negative and far from zero by several standard deviations. It is worth noting that a pure relativistic reflection model was the model employed in~\cite{nustar} to measure the black hole spin assuming the Kerr metric. Note also that our measurements of the model parameters for the Kerr case are consistent with those reported in~\cite{nustar} even if there are some minor differences between the two models [\cite{nustar} use {\sc reflionx} as non-relativistic reflection model while here we use {\sc xillver}]. In particular, the two spin measurements are consistent at 1-$\sigma$ when we assume the Kerr metric.

We add a non-relativistic reflection component and the total model becomes {\sc tbabs*(diskbb + relxill\_nk + xillver)}. Such a component can be easily generated, for instance, by some outflow from the accretion disk. First, we model the emissivity profile with a simple power law: in model~3 we assume the Kerr metric, in model~3$a$ we have $\alpha_{13}$ free and $\alpha_{22}=0$, and in model~3$b$ we have $\alpha_{13}=0$ and $\alpha_{22}$ is free. The best-fit values for the Kerr model are reported in Tab.~\ref{t-fit-kerr}. As shown in Tab.~\ref{t-fit-nustar} and in Fig.~\ref{f-plots2-nustar}, the measurements of $\alpha_{13}$ and $\alpha_{22}$ are now both consistent with the Kerr solution.

\begin{figure*}[t]
\begin{center}
\includegraphics[width=8.5cm,trim={1cm 0 0 0},clip]{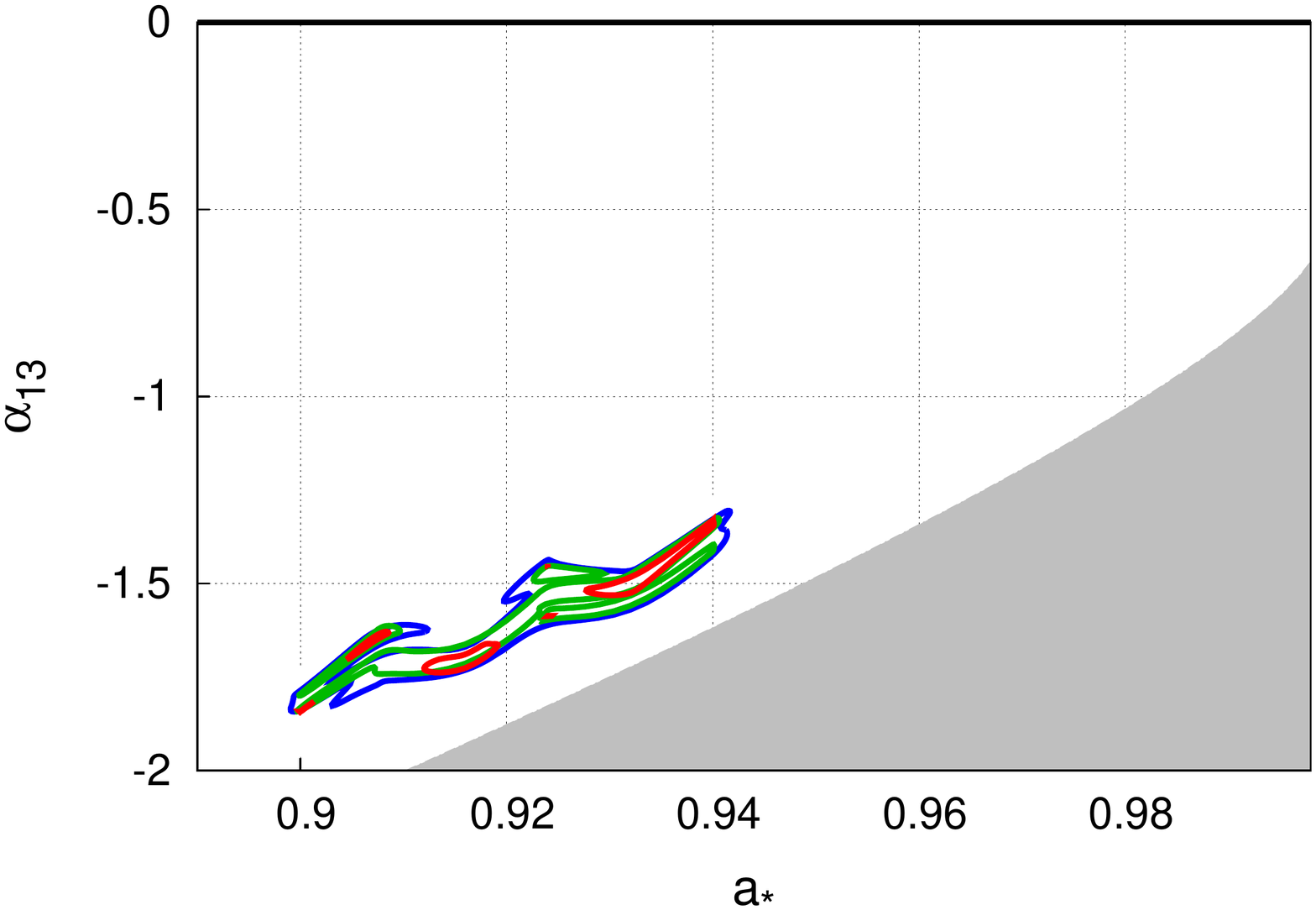}
\includegraphics[width=8.5cm,trim={1cm 0 0 0},clip]{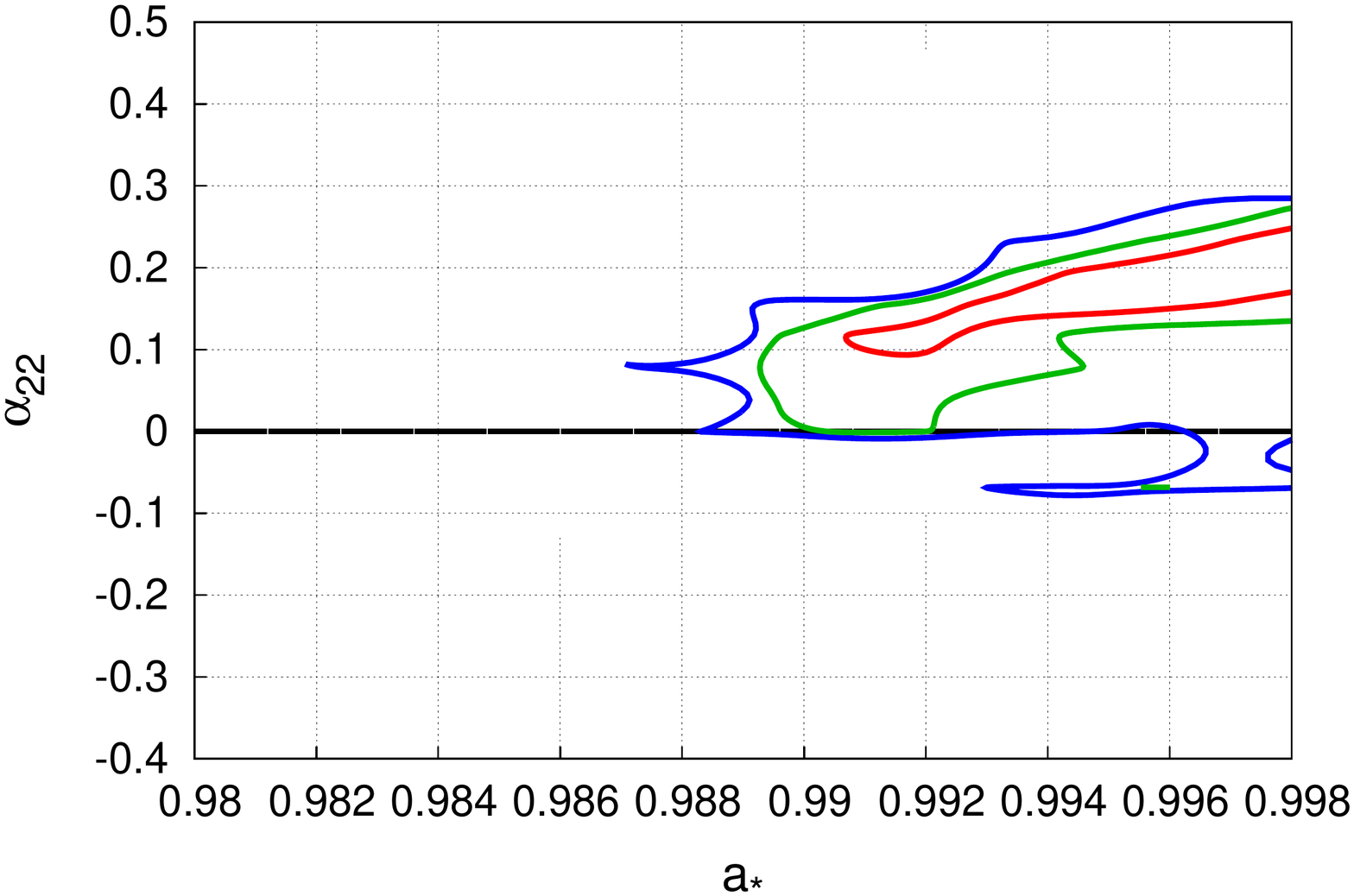}
\end{center}
\vspace{-1.3cm}
\caption{Left panel: Constraints on the spin parameter $a_*$ and on the Johannsen deformation parameter $\alpha_{13}$ according to model~3$a'$. Right panel: Constraints on the spin parameter $a_*$ and on the Johannsen deformation parameter $\alpha_{22}$ according to model~3$b'$. The red, green, and blue lines indicate, respectively, the 68\%, 90\%, and 99\% confidence level contours for two relevant parameters. The solid black line marks the Kerr solution. The grayed region is ignored in our study because it does not meet the condition in Eq.~(\ref{eq-constraints}). \label{f-plots3-nustar}}
\end{figure*}

Second, we employ a broken power law to describe the emissivity profile: in model~3$'$ we assume the Kerr metric, in model~3$a'$ we have $\alpha_{13}$ free and $\alpha_{22}=0$, and in model~3$b'$ we have $\alpha_{13}=0$ and $\alpha_{22}$ is free. As we can see from Fig.~\ref{f-plots3-nustar}, the measurement of $\alpha_{22}$ is consistent with the Kerr solution. However, the measurement of $\alpha_{13}$ is very far from zero: the difference of $\chi^2$ between models 3$'$ and 3$a'$ is $\Delta\chi^2 = 29$. From Tab.~\ref{t-fit-nustar}, we see that the inner emissivity index is lower than the outer one for models~3$a'$ and $3b'$. In particular, we find $q_{\rm out}$ very high in both cases. This simply means that the fit prefer a disk with a relatively constant emissivity near the inner edge and then a very weak emissivity at larger radii.

Tab.~\ref{t-models} lists the main models discussed above and employed in our spectral analysis, as well as the corresponding properties.

\begin{table*}
\centering
\vspace{0.5cm}
\begin{tabular}{lccc}
\hline\hline
Model & 2 & 3 & 3$'$ \\
\hline
{\sc tbabs} && \\
$N_{\rm H} / 10^{22}$ cm$^{-2}$ & $8.93^{+0.31}_{-0.06}$ & $7.1_{-0.6}^{+0.6}$ & $8.1^{+0.3}_{-0.4}$ \\
\hline
{\sc diskbb} && \\
$T_{\rm in}$ [keV] & $0.4205^{+0.0011}_{-0.0014}$ & $0.427^{+0.029}_{-0.025}$ & $0.400^{+0.016}_{-0.011}$ \\
\hline
{\sc relxill\_nk} && \\
$q_{\rm in}$ & $> 9.8$ & $4.7^{+4.3}_{-1.2}$ & $< 3.3$ \\
$q_{\rm out}$ & $= q_{\rm in}$ & $= q_{\rm in}$ & $> 9.4$ \\
$R_{\rm br}$~$[M]$ & -- & -- & $1.71^{+0.06}_{-0.07}$ \\
$i$ [deg] & $75.59^{+0.23}_{-0.16}$ & $64^{+9}_{-3}$ & $75.9^{+0.7}_{-0.9}$ \\
$a_*$ & $0.9875^{+0.0006}_{-0.0056}$ & $0.967^{+0.012}_{-0.025}$ & $> 0.989$ \\
$\alpha_{13}$ & $0^\star$ & $0^\star$ & $0^\star$ \\
$\alpha_{22}$ & $0^\star$ & $0^\star$ & $0^\star$ \\
$\log\xi$ & $3.025^{+0.028}_{-0.014}$ & $3.47^{+0.25}_{-0.16}$ & $3.04^{+0.04}_{-0.03}$ \\
$A_{\rm Fe}$ & $0.907^{+0.021}_{-0.081}$ & $1.1^{+0.8}_{-0.3}$ & $0.67^{+0.07}_{-0.07}$ \\
$\Gamma$ & $2.080^{+0.004}_{-0.004}$ & $1.89^{+0.05}_{-0.08}$ & $2.13^{+0.03}_{-0.05}$ \\
$E_{\rm cut}$ [keV] & $60.6^{+0.5}_{-1.0}$ & $47^{+6}_{-5}$ & $69^{+5}_{-8}$ \\
$R_{\rm f}$ & $0.228^{+0.006}_{-0.011}$ & $0.17^{+0.07}_{-0.03}$ & $0.27^{+0.03}_{-0.04}$ \\
\hline
{\sc xillver} && \\
$\log\xi$ & -- & $2.80^{+0.14}_{-0.09}$ & $2.30^{+0.09}_{-0.15}$ \\
\hline
$\chi^2/\nu$ & $2630.40/2388 \quad$ & $\quad 2546.33/2386 \quad$ & $\quad 2537.82/2384 \quad$ \\
& =1.10151 & =1.06719 & =1.06452 \\
\hline\hline
\end{tabular}
\vspace{0.2cm}
\caption{Summary of the best-fit values for models~2, 3, and 3$'$ (Kerr spacetime with $\alpha_{13} = \alpha_{22} = 0$). The reported uncertainties correspond to the 90\% confidence level for one relevant parameter. $^\star$ indicates that the parameter is frozen. \label{t-fit-kerr}}
\end{table*}

\begin{table*}
\centering
\vspace{0.5cm}
\begin{tabular}{lcccccc}
\hline\hline
Model & 2$a$ & 2$b$ & 3$a$ & 3$b$ & 3$a'$ & 3$b'$ \\
\hline
{\sc tbabs} && \\
$N_{\rm H} / 10^{22}$ cm$^{-2}$ & $7.4^{+0.4}_{-0.4}$ & $8.69_{-0.36}^{+0.20}$ & $7.0^{+0.6}_{-0.9}$ & $7.1^{+0.6}_{-0.6}$ & $8.66^{+0.13}_{-0.12}$ & $8.1^{+0.5}_{-0.3}$ \\
\hline
{\sc diskbb} && \\
$T_{\rm in}$ [keV] & $0.418^{+0.018}_{-0.020}$ & $0.425^{+0.010}_{-0.016}$ & $0.43^{+0.03}_{-0.03}$ & $0.427^{+0.023}_{-0.022}$ & $0.3537^{+0.0012}_{-0.0213}$ & $0.402^{+0.014}_{-0.018}$ \\
\hline
{\sc relxill\_nk} && \\
$q_{\rm in}$ & $5.7^{+1.4}_{-0.9}$ & $> 9.7$ & $3.3^{+2.6}_{-0.9}$ & $4.7^{+3.0}_{-2.0}$ & $0.2^{+1.2}$ & $4.2^{+2.1}_{-2.1}$ \\
$q_{\rm out}$ & $= q_{\rm in}$ & $= q_{\rm in}$ & $= q_{\rm in}$ & $= q_{\rm in}$ & $7.3_{-1.5}$ & $> 8$ \\
$R_{\rm br}$~$[M]$ & -- & -- & -- & -- & $1.85^{+0.35}_{-0.04}$ & $1.40^{+0.14}_{-0.06}$ \\
$i$ [deg] & $64.9^{+0.4}_{-0.5}$ & $74.5^{+0.7}_{-0.4}$ & $62.2^{+1.2}_{-2.3}$ & $64^{+6}_{-3}$ & $68.8^{+0.6}_{-0.7}$ & $77.8^{+0.9}_{-5.5}$ \\
$a_*$ & $> 0.993$ & $> 0.995$ & $0.989_{-0.010}$ & $0.971^{+0.021}_{-0.071}$ & $0.913^{+0.014}_{-0.009}$ & $> 0.990$ \\
$\alpha_{13}$ & $-0.50^{+0.02}_{-0.01}$ & $0^\star$ & $0.2^{+0.1}_{-1.7}$ & $0^\star$ & $-1.7^{+0.4}_{-0.1}$ & $0^\star$ \\
$\alpha_{22}$ & $0^\star$ & $-0.13^{+0.05}_{-0.01}$ & $0^\star$ & $0.0^{+0.7}_{-0.2}$ & $0^\star$ & $0.21^{+0.04}_{-0.12}$ \\
$\log\xi$ & $2.96^{+0.05}_{-0.11}$ & $3.04^{+0.03}_{-0.07}$ & $3.51^{+0.14}_{-0.08}$ & $3.47^{+0.21}_{-0.17}$ & $2.85^{+0.07}_{-0.03}$ & $3.05^{+0.04}_{-0.04}$ \\
$A_{\rm Fe}$ & $1.4^{+0.4}_{-0.4}$ & $0.97^{+0.18}_{-0.23}$ & $1.01^{+0.60}_{-0.16}$ & $1.07^{+1.03}_{-0.17}$ & $0.560^{+0.051}_{-0.017}$ & $0.67_{-0.05}^{+0.12}$ \\
$\Gamma$ & $2.01^{+0.04}_{-0.03}$ & $2.044^{+0.015}_{-0.011}$ & $1.889^{+0.023}_{-0.030}$ & $1.89^{+0.05}_{-0.07}$ & $2.261^{+0.021}_{-0.034}$ & $2.13^{+0.05}_{-0.04}$ \\
$E_{\rm cut}$ [keV] & $51.7^{+2.2}_{-2.9}$ & $56.7^{+6.5}_{-1.6}$ & $47^{+4}_{-3}$ & $47^{+5}_{-6}$ & $87.2^{+3.3}_{-1.8}$ & $69^{+7}_{-7}$ \\
$R_{\rm f}$ & $0.209^{+0.015}_{-0.017}$ & $0.202^{+0.066}_{-0.013}$ & $0.16^{+0.07}_{-0.03}$ & $0.17^{+0.06}_{-0.03}$ & $0.345^{+0.024}_{-0.014}$ & $0.273^{+0.021}_{-0.021}$ \\
\hline
{\sc xillver} && \\
$\log\xi$ & -- & -- & $2.84^{+0.15}_{-0.08}$ & $2.80^{+0.13}_{-0.09}$ & $3.10^{+0.08}_{-0.20}$ & $2.29^{+0.09}_{-0.16}$ \\
\hline
$\chi^2/\nu$ & $2564.57/2387 \quad$ & $\quad 2622.33/2387 \quad$ & $\quad 2545.12/2385 \quad$ & $\quad 2546.35/2385 \quad$ & $\quad 2508.85/2383 \quad$ & $\quad 2535.54/2383 \quad$ \\
& =1.07439 & =1.09859 & =1.06714 & =1.06765 & 1.05281 & 1.06401 \\
\hline\hline
\end{tabular}
\vspace{0.2cm}
\caption{Summary of the best-fit values for models~2$a$, 2$b$, 3$a$, 3$b$, 3$a'$, and 3$b'$. The reported uncertainties correspond to the 90\% confidence level for one relevant parameter. $^\star$ indicates that the parameter is frozen. \label{t-fit-nustar}}
\end{table*}

\begin{table*}
\centering
\vspace{0.5cm}
\begin{tabular}{lcc}
\hline\hline
Xspec model & $\qquad$ Model $\qquad$ & Description \\
\hline
{\sc tbabs*cutoffpl} & 0 & \\
\hline
{\sc tbabs*relxill\_nk} & 1 & Kerr, $q_{\rm out} = q_{\rm in}$ \\
& 1$a$ & $\alpha_{13}$ free, $q_{\rm out} = q_{\rm in}$ \\
& 1$b$ & $\alpha_{22}$ free, $q_{\rm out} = q_{\rm in}$ \\
\hline
{\sc tbabs*(diskbb + relxill\_nk)} & 2 & Kerr, $q_{\rm out} = q_{\rm in}$ \\
& 2$a$ & $\alpha_{13}$ free, $q_{\rm out} = q_{\rm in}$ \\
& 2$b$ & $\alpha_{22}$ free, $q_{\rm out} = q_{\rm in}$ \\
\hline
{\sc tbabs*(diskbb + relxill\_nk + xillver)} & 3 & Kerr, $q_{\rm out} = q_{\rm in}$ \\
& 3$a$ & $\alpha_{13}$ free, $q_{\rm out} = q_{\rm in}$ \\
& 3$b$ & $\alpha_{22}$ free, $q_{\rm out} = q_{\rm in}$ \\
\hline
{\sc tbabs*(diskbb + relxill\_nk + xillver)} & 3$'$ & Kerr, $q_{\rm out}$ free \\
& 3$a'$ & $\alpha_{13}$ free, $q_{\rm out}$ free \\
& 3$b'$ & $\alpha_{22}$ free, $q_{\rm out}$ free \\
\hline
\hline\hline
\end{tabular}
\vspace{0.2cm}
\caption{List of the main models employed in our spectral analysis. \label{t-models}}
\end{table*}

\section{Discussion and conclusions \label{s-dis}}

When we add a non-relativistic reflection component to the model, the quality of the fit improves and we can argue that such a non-relativistic reflection component is indeed necessary. It is thus perfectly understandable that we do not recover the Kerr solution in models~2$a$, 2$b$, and their variants with a different emissivity profile. We are missing an important component in the spectrum and we cannot pretend to test the Kerr metric with GRS~1915+105. The take-away message is that the choice of the correct model can be very important in these kinds of tests.

With the non-relativistic reflection component in the model, the measurement of $\alpha_{13}$ and $\alpha_{22}$ turn out to be very sensitive to the choice of the shape of the emissivity profile. In models~3$a$ and 3$b$, we recover the Kerr solution, but the constraints on $\alpha_{13}$ and $\alpha_{22}$ are weak. In model~3$a'$, we do not recover the Kerr solution at a high confidence level. In model~3$b'$, we recover the Kerr solution and the constraint is strong. Note that we cannot say that the correct astrophysical model is the one in which we recover the Kerr metric, because this would be equivalent to saying that we want to test the astrophysical model and we assume the Kerr metric. We have thus to figure out how we can separately test the metric and the astrophysical model.

We note that we are not able to fit well the high energy part of the spectrum, see Fig.~\ref{f-ratio-nustar}. Since the cut-off energies that we obtain are very low, the corona should be relatively cool. In such a case, the difference between a simple cut-off power law and a proper comptonization model may be important. We have thus repeated our analyses by replacing {\sc relxill\_nk} with {\sc relxillCp\_nk} in our models~\cite{noi0b}. However, we have obtained worse fits. We have also tried other solutions, like adding an extra power law component to describe the possible emission from the jet, but still we are not able to improve the quality of the fit at high energies.

In our previous analyses of stellar-mass and supermassive black holes with {\sc relxill\_nk}, we had never found similar problems. First, we were able to easily recover vanishing values of $\alpha_{13}$ and $\alpha_{22}$. Second, the choice between power law and broken power law for the description of the intensity profile had not such a strong impact on the final estimate of $\alpha_{13}$ and $\alpha_{22}$. More specifically, we usually found that a power law or a broken power law could provide somewhat different but consistent results. On the contrary, imposing an {\it ad hoc} emissivity profile (i.e. without fitting the emissivity indices and the breaking radius), we obtained non-vanishing values of $\alpha_{13}$ and $\alpha_{22}$. With such results, we argued that the emissivity profile is important to correctly model the spectrum of the source, but that it is possible to separately measure the deformation parameters and the parameters related to the emissivity profile. The case of GRS~1915+105 seems to be different.

It is likely that the spectrum of GRS~1915+105 is more difficult to model. While the spectrum may indeed be described by a thermal component from the disk and relativistic and non-relativistic reflection components, {\sc relxill\_nk} and {\sc xillver} may not be able to properly describe these components. Both models have indeed a number of simplifications. If the theoretical model does not properly describe the observed spectrum, the fit tries to absorb such a discrepancy with incorrect values of the parameters.

A crucial assumption in our reflection model {\sc relxill\_nk} is that the accretion disk is thin and the inner edge is at the ISCO radius. If the actual accretion disk around the black hole does not meet these conditions, we can have systematic uncertainties that can mimic a non-vanishing deformation parameter. However, we do not think this is the reason for our results for GRS~1915+105. For the \textsl{NuSTAR} observation of 2012, the Eddington-scaled accretion luminosity of the source is $0.23 \pm 0.04$~\cite{nustar}, which is in the 0.05-0.30 range required to have thin disks~\cite{thin1,thin2}. On the contrary, in other works we have obtained quite stringent constraints on $\alpha_{13}$ and $\alpha_{22}$ from supermassive black holes that more likely accrete above 30\% of their Eddington limit.

The current version of {\sc xillver} is appropriate for the description of ``cold'' disks, because it neglects the contribution in the reflection spectrum from the X-ray photons emitted by the disk itself. This may explain our difficulties in recovering the Kerr metric. In the case of GS~1354--645, the fit did not need any thermal component, which means that the disk's temperature is lower than the one in the \textsl{NuSTAR} observation of GRS~1915+105. In the case of supermassive black holes, the temperature of the disk is a few orders of magnitude lower, so {\sc xillver} is appropriate. We note that we are meeting a similar problem in recovering the Kerr solution in a work in preparation on Cygnus~X-1, where we analyze some \textsl{NuSTAR} observations in which the source is in the soft state.

Lastly, we note that in all our models we have always assumed that the inner edge of the accretion disk is at the ISCO radius, i.e. $R_{\rm in} = R_{\rm ISCO}$. This is quite a common choice, both when we try to measure the black hole spin assuming the Kerr metric and when we want to test the Kerr hypothesis. If we do not do so, in general it is impossible to measure the spin and the deformation parameter because of the strong degeneracy of the latter with $R_{\rm in}$. However, this is not the case when the black hole is rotating very fast, with a spin parameter close to the maximum value allowed by the model, such as for GRS~1915+105. As we have already shown in~\cite{noi4} for GS~1354--645 and the Johannsen metric, leaving $R_{\rm in}$ as a free parameter has quite a negligible impact on the estimate of the other model parameters just because $R_{\rm in}$ is already constrained to be close to minimum value allowed by the model. Note that such a conclusion cannot be automatically extended to any metric.

The possibility of performing precise tests of the Kerr metric using X-ray reflection spectroscopy, which is our long-term goal, requires having a theoretical model good enough such that the systematic uncertainties are subdominant with respect to the statistical ones. In such a case, it is important to be able to select the right candidates, namely observations of black holes that can be well described by the available theoretical model. This will become of crucial importance with the next generation of X-ray missions, like \textsl{eXTP} and \textsl{Athena}, that promise to provide unprecedented high quality data~\cite{extp}.


{\bf Acknowledgments --}
This work was supported by the National Natural Science Foundation of China (NSFC), Grant No.~U1531117, and Fudan University, Grant No.~IDH1512060. Y.Z. also acknowledges the support from the Fudan Undergraduate Research Opportunities Program (FDUROP). A.B.A. also acknowledges the support from the Shanghai Government Scholarship (SGS). J.A.G. acknowledges support from the Alexander von Humboldt Foundation. S.N. acknowledges support from the Excellence Initiative at Eberhard-Karls Universit\"at T\"ubingen.


\end{document}